\documentclass[twocolumn]{aastex702}  

\usepackage{amsmath}
\usepackage{graphicx}
\usepackage{threeparttable}
\usepackage{txfonts}
\hypersetup{allcolors=blue}

\newcommand{\gtsima}{$\; \buildrel > \over \sim \;$}
\newcommand{\ltsima}{$\; \buildrel < \over \sim \;$}
\newcommand{\prosima}{$\; \buildrel \propto \over \sim \;$}
\newcommand{\gsim}{\lower.5ex\hbox{\consistegtsima}}
\newcommand{\lsim}{\lower.5ex\hbox{\ltsima}}
\newcommand{\simgt}{\lower.5ex\hbox{\gtsima}}
\newcommand{\simlt}{\lower.5ex\hbox{\ltsima}}
\newcommand{\simpr}{\lower.5ex\hbox{\prosima}}

\newcommand\iona[2]{#1$\;${\scshape{#2}}}% iona, i.e. Mg II = \iona{Mg}{ii}

\renewcommand{\footnotesize}{\small}

\begin{document}  

\title{Revisiting the Claim for a Direct-Collapse Black Hole in UHZ1 at \textit{z} = 10.05}

\author[0000-0002-4436-6923]{Fan Zou}
\affiliation{Department of Astronomy, University of Michigan, 1085 S University, Ann Arbor, MI 48109, USA}
\email[show]{fanzou01@gmail.com}

\author[0000-0001-5802-6041]{Elena Gallo}
\affiliation{Department of Astronomy, University of Michigan, 1085 S University, Ann Arbor, MI 48109, USA}
\email[show]{egallo@umich.edu}

\author[0009-0000-8524-8344]{Zihao Zuo}
\affiliation{Department of Astronomy, University of Michigan, 1085 S University, Ann Arbor, MI 48109, USA}
\email{zuo@umich.edu}

\author[0000-0002-2397-206X]{Edmund Hodges-Kluck}
\affiliation{X-ray Astrophysics Laboratory, NASA/GSFC, Greenbelt, MD 20771, USA}
\email{Edmund.hodges-kluck@nasa.gov}

\author[0000-0002-5678-1008]{Dieu D. Nguyen}
\affiliation{Department of Astronomy, University of Michigan, 1085 S University, Ann Arbor, MI 48109, USA}
\email{dieun@umich.edu}

\author[0000-0002-4140-1367]{Guido Roberts-Borsani}
\affiliation{Department of Physics \& Astronomy, University College London, London, WC1E 6BT, UK}
\email{g.robertsborsani@ucl.ac.uk}

\author[0000-0002-6336-3293]{Piero Madau}
\affiliation{Department of Astronomy \& Astrophysics, University of California, 1156 High Street, Santa Cruz, CA 95064, USA}
\affiliation{Dipartimento di Fisica ``G. Occhialini," Università degli Studi di Milano-Bicocca, Piazza della Scienza 3, I-20126 Milano, Italy}
\email{pmadau@ucsc.edu}

\author[0000-0001-9879-7780]{Fabio Pacucci}
\affiliation{Center for Astrophysics, Harvard \& Smithsonian, 60 Garden St, Cambridge, MA 02138, USA}
\affiliation{Black Hole Initiative, Harvard University, 20 Garden St, Cambridge, MA 02138, USA}
\email{fabio.pacucci@cfa.harvard.edu}

\author[0000-0003-0248-5470]{Anil C. Seth}
\affiliation{Department of Physics and Astronomy, University of Utah, USA}
\email{aseth@astro.utah.edu}

\author[0000-0002-8460-0390]{Tommaso Treu}
\affiliation{Physics and Astronomy Department, University of California, Los Angeles, CA 90095, USA}
\email{tt@astro.ucla.edu}

\author[0009-0003-1260-4143]{Shouyi Wang}
\email{sywang0302@gmail.com}
\affiliation{Department of Astronomy, School of Physics, Peking University, Beijing 100871, People's Republic of China}
\affiliation{Kavli Institute for Astronomy and Astrophysics, Peking University, Beijing 100871, People's Republic of China}

\begin{abstract}
We reassess the direct collapse black hole (DCBH) interpretation of UHZ1 
(UNCOVER-26185), a gravitationally lensed galaxy at $z_\mathrm{spec}=10.054$. 
That interpretation rests on a hard ($2$--$7$~keV) X-ray excess detected with 
Chandra, attributed to a Compton-thick AGN with an inferred 
$2$--$10$~keV luminosity of $L_\mathrm{X,int}\sim10^{46}~\mathrm{erg~s^{-1}}$ 
\citep{Bogdan2024}. The resulting extreme X-ray to rest-frame optical--IR ratio 
was taken as the hallmark signature of an ``outsize black hole galaxy'' at 
cosmic dawn. We analyse the full 2.2~Ms Chandra imaging 
dataset---including 0.95~Ms of unpublished observations---and 
present new JWST/MIRI photometry at $\lambda_\mathrm{obs}>5~\mu\mathrm{m}$. 
Across the full range of plausible Chandra data reductions, the 
$2$--$7$~keV excess at the position of UHZ1 reaches a significance of only 
$2.0$--$2.9\sigma$; the originally reported $4.2$--$4.4\sigma$ detection is 
sensitive to the specific astrometric alignment adopted and is not robustly 
reproducible. Moreover, the hard X-ray signal does not grow with the 
additional exposure, contrary to expectations for a steady source, indicating 
that any excess is not persistent. UHZ1 is also undetected in all nine MIRI 
imaging bands. Fitting red/obscured AGN SED templates to the tightest MIRI 
upper limit, we constrain the bolometric luminosity of any buried AGN to 
$L_\mathrm{bol}<1.3\times10^{45}~\mathrm{erg~s^{-1}}$. These conclusions are 
further supported by independent JWST spectroscopy 
\citep{Alvarez-Marquez2026}, which reveals no AGN signatures in the rest-frame 
UV or optical. Taken together, the multiwavelength data paint a consistent 
picture of UHZ1 as a low-mass, metal-poor, star-forming galaxy in the early 
Universe, with no compelling evidence for a luminous obscured AGN, regardless 
of its proposed formation channel.
\end{abstract}


%%%%%%%%%%%%%%%%%%%%%
\section{Introduction}
\label{sec:intro}

The origin of supermassive black holes (SMBHs) is one of the most
pressing unsolved problems at the intersection of high-energy
astrophysics and galaxy formation. Over three hundred quasars powered
by SMBHs with masses $M_{\rm BH}\gtrsim 10^9\,M_\odot$ have now been
spectroscopically confirmed at $z\gtrsim 6$
\citep{Fan2001,Fan2003,Fan2023,Mortlock2011,Wu2015,Banados2018,Wang2021},
implying that billion-solar-mass black holes were already in place less
than one billion years after the Big Bang. Growing such objects from stellar-mass remnants ($M_{\rm seed}\sim 10$--$10^2\,M_\odot$) is challenging and, under standard radiatively efficient growth, requires nearly continuous Eddington-rate accretion for several hundred Myr \citep{Madau2001,Haiman2004,Volonteri2010,Inayoshi2020,Pacucci2022}. This has motivated both theoretical work on more massive initial seeds \citep{Rees1984,Volonteri2012,Natarajan2014} and alternative rapid-growth scenarios for light seeds \citep{Madau2014}.

The most widely discussed alternative is the direct collapse
scenario, in which primordial gas in atomic-cooling haloes
($T_{\rm vir}\gtrsim 10^4$\,K) avoids fragmentation into stars when
H$_2$ cooling is suppressed by a strong Lyman--Werner radiation field,
and instead collapses to form a direct collapse black hole
(DCBH) seed with $M_{\rm seed}\sim 10^4$--$10^6\,M_\odot$
\citep{Bromm2003,Begelman2006,Begelman2008,Lodato2006,
Volonteri2008,Inayoshi2020}.
Starting from such a heavy seed would largely alleviate the timing problem, enabling growth to $\simeq 10^9~M_\odot$ by $z\sim 6$ under more moderate accretion histories.

A key prediction of this
scenario is that, during the rapid early growth phase, the BH mass can
be comparable to or exceed the stellar mass of the host galaxy---the
so-called ``outsize black hole galaxy"
\citep[OBG;][]{Natarajan2017,Scoggins2023,Scoggins2024} phase---in stark contrast to the
$M_{\rm BH}/M_\star \sim 10^{-3}$ relation in the local Universe
\citep{Kormendy2013}. This extreme mass ratio manifests
observationally as a very high ratio of rest-frame X-ray to optical
flux: the accreting heavy seed dominates the bolometric output while
the stellar mass of the host remains low, producing a source that is
luminous in hard X-rays yet faint and red in the UV--optical
\citep{Pacucci2015,Pacucci2016,Natarajan2017}. This distinctive multiwavelength
signature is, in principle, what uniquely identifies a DCBH candidate
at high redshift and distinguishes it from the more ordinary
over-massive BHs increasingly reported at $z\sim 4$--7 by JWST
\citep{Pacucci2023,Harikane2023,Maiolino2024}. However, it is worth noting that the observed \mbox{X-rays} may also be suppressed. During the early phase of the DCBH accretion, the \mbox{X-ray} emission may be significantly absorbed at a Compton-thick level \citep{Pacucci2015, Pacucci2026}. The \mbox{X-rays} may also be intrinsically weak due to, e.g., super-Eddington accretions, which have been recently proposed to explain the \mbox{X-ray} weakness of high-redshift AGNs (e.g., \citealt{Madau2024, Pacucci2024, Trinca2026}). The level of any such \mbox{X-ray} weakness is therefore model- and phase-dependent. Conversely, a robustly high X-ray-to-optical flux ratio would remain difficult to explain without an over-massive, rapidly accreting BH, and would thus still constitute distinctive evidence for the OBG/DCBH scenario.

Against this backdrop, the discovery of UHZ1 (UNCOVER-26185)\footnote{This UNCOVER ID is quoted here to be consistent with the literature \citep{Fujimoto2024, Alvarez-Marquez2026}. However, the UNCOVER source IDs are not persistent identifiers. UHZ1 has different IDs of 27025 and 38201 in the UNCOVER DR2 \citep{Weaver2024} and DR3 \citep{Suess2024} photometric catalogs, respectively.} at $z \simeq 10.05$ was immediately recognized as potentially decisive.
The source was first identified as a high-redshift candidate in deep
NIRCam imaging of the Frontier Fields cluster Abell~2744
\citep{Castellano2022,Castellano2023,Bezanson2024}. Its redshift was
subsequently confirmed spectroscopically at $z_{\rm spec} = 10.054$ via
JWST/NIRSpec Prism and JWST/MIRI LRS observations, which
also yielded a stellar mass $M_\star = (1.7 \pm 0.3) \times 10^8\,M_\odot$
and a low metallicity $Z = (0.04 \pm 0.01)\,Z_\odot$
\citep{Goulding2023,Alvarez-Marquez2026}. 

Gravitational lensing by Abell~2744
magnifies the source by a factor $\mu = 3.71_{-0.23}^{+0.23}$
\citep{Bergamini2023}, bringing it within reach of X-ray observation.
Leveraging 1.25\,Ms of archival Chandra/ACIS imaging of the
Abell~2744 field, \citet{Bogdan2024} reported a $4.2$--$4.4\,\sigma$
excess of hard X-ray counts (2--7\,keV; rest-frame 22--88\,keV)
spatially coincident with UHZ1. The detection rests on only $\sim$21
net counts above a background of $\sim$21.4 counts, the latter
dominated by the hot ($kT\sim 11$\,keV) intra-cluster medium (ICM) of
Abell~2744. The source is undetected in the soft band, consistent with
extreme absorption. 
Spectral fitting with a Compton-thick AGN model \citep{Yaqoob2012}
yielded a best-fit column density $N_\mathrm{H}$ of $8\times10^{24}~\mathrm{cm^{-2}}$ and $N_\mathrm{H}>1\times10^{24}~\mathrm{cm^{-2}}$ at a $1\sigma$ level, though the authors acknowledged severe
degeneracy between $N_{\rm H}$ and $L_\mathrm{X}$ owing to the limited photon
statistics. Adopting the lower $1\sigma$ bound on $N_{\rm H}$ and
assuming the resulting bolometric luminosity to be Eddington-limited
yields $M_{\rm BH} \sim 10^{7}$--$10^{8}\,M_\odot$, comparable to
the host galaxy stellar mass (adopting instead the best-fit $N_{\rm H}$ would
yield a black hole mass exceeding the entire host galaxy). The authors
therefore interpreted UHZ1 as the first observational evidence for
heavy-seed DCBH formation \citep{Natarajan2024}.

This interpretation rested on assumptions that were already sources of
uncertainty at the time of publication. The published X-ray signal is
below the $5\sigma$ threshold warranted for extraordinary claims---and the BH mass inference
is the product of a long chain of model-dependent steps. The
NIRSpec Prism spectrum \citep{Goulding2023} showed neither broad lines nor
high-ionization UV emission, attributed to extreme obscuration, but
the strongest rest-frame optical diagnostics (H$\alpha$, H$\beta$,
[\iona{O}{iii}]) lay outside the NIRSpec wavelength coverage at
$z\simeq 10$, preventing a direct test.

Independent confirmation was therefore urgently needed. The MIRI
instrument on JWST provides a uniquely powerful diagnostic: at
$z\simeq 10$, observed wavelengths $5\lesssim\lambda_{\rm obs}\lesssim
25\,\mu$m sample the rest-frame $0.4$--$2.2\,\mu$m window,
encompassing both the classical rest-optical emission lines used in AGN
diagnostics \citep[BPT;][]{Baldwin1981,Kewley2001,Kauffmann2003} and
the rest-frame $\sim$1--2\,$\mu$m near-IR continuum where thermal
emission from a hot AGN torus becomes visible \citep{Hickox2018,Lyu2022}. This
torus emission acts as a calorimeter for the buried AGN: regardless of
the X-ray column density, an AGN with $L_{\rm bol}\sim
10^{45}$\,erg\,s$^{-1}$ must heat its surrounding dust and re-radiate
a substantial fraction of that power in the near-IR, making mid-IR
photometry a genuinely independent probe of the AGN hypothesis. Hot
dusty tori have been directly detected with JWST/MIRI out to $z\sim 7$
\citep{Bosman2024}, demonstrating the feasibility of this approach. Nevertheless, the hot dust emission may be suppressed if there is no dust or the obscuring dust is far from the AGN and is thus cooler. These possibilities will also be discussed in detail in Section~\ref{sec: discussion}.

Recent JWST observations have provided exactly these independent
constraints. \citet{Alvarez-Marquez2026} presented deep (13.9\,hr)
JWST/MIRI LRS spectroscopy of UHZ1 as part of the PRImordial galaxy
Survey with MIRI Spectroscopy (PRISMS; PID~8051), detecting
H$\beta$+[\iona{O}{iii}]$\lambda\lambda$4960,5008 and H$\alpha$ at
$10\sigma$ and $8\sigma$, respectively. The inferred nebular properties
are inconsistent with a luminous obscured AGN: the UV-to-optical SED
is well described by a low-dust stellar population with emission-line
ratios placing the source in the low-metallicity star-forming locus of
standard diagnostic diagrams, with no AGN ionizing continuum required.
These results are consistent with an independent analysis by
\citet{Fujimoto2024}.

UHZ1 is therefore at the center of a growing multi-wavelength tension
with direct implications for the DCBH interpretation and, more
broadly, for the reliability of X-ray--based BH mass estimates at
cosmic dawn. In this paper, we present a comprehensive reassessment
using two new datasets. First, we reprocess the full Chandra
dataset of \citet{Bogdan2024} and incorporate an additional 0.95\,Ms of
exposure obtained between 2023-05-24 and 2024-05-29, applying a homogeneous reduction and careful local treatment of
the cluster background. Second, we present
new JWST/MIRI imaging extending to $\lambda_{\rm obs} > 5\,\mu$m,
providing the first direct constraints on the rest-frame near-IR
continuum. 

The remainder of this paper is structured as follows. The multiwavelength dataset is described in Section~\ref{sec: data}, where Section~\ref{sec: cxo} presents the Chandra reanalyses, and Section~\ref{sec: miri} presents the MIRI imaging. Implications for the DCBH scenario are discussed in Section~\ref{sec: discussion}. Throughout we assume a flat $\Lambda$CDM cosmology with $H_0 = 67.7$\,km\,s$^{-1}$\,Mpc$^{-1}$ and $\Omega_M = 0.31$ \citep{PlanckCollaboration2020}.

\section{Observations, Data Reduction, and Analysis}
\label{sec: data}

\subsection{Chandra/ACIS}
\label{sec: cxo}

We (re-)analyse the Chandra/ACIS imaging at the sky
position of UHZ1 to independently assess the significance of the hard-band
excess reported by \citet{Bogdan2024}. Our approach proceeds in four steps:
(\textit{i})~homogeneous data reduction of both the original and extended
datasets (Section~\ref{sec: cxo_reduction});
(\textit{ii})~aperture photometry with careful treatment of the structured
ICM background and a direct comparison with the published
detection (Section~\ref{sec: cxo_photometry});
(\textit{iii})~a Monte Carlo assessment of the sensitivity of the detection
significance to astrometric registration choices
(Section~\ref{sec: cxo_astro}); and
(\textit{iv})~a temporal analysis of the hard-band signal across individual
epochs to test the hypothesis of a steady source
(Section~\ref{sec: cxo_lc}).

\begin{table*}
\caption{Chandra observations of the field of view of UHZ1 (Abell 2744)}
\label{tbl: chandra_obs}
\centering
\begin{threeparttable}
\begin{tabular}{cccc|cccc|cccc}
\hline\hline
ObsID & Date & $T_\mathrm{exp}$ & $T_\mathrm{exp}^\mathrm{clean}$ & ObsID & Date & $T_\mathrm{exp}$ & $T_\mathrm{exp}^\mathrm{clean}$ & ObsID & Date & $T_\mathrm{exp}$ & $T_\mathrm{exp}^\mathrm{clean}$\\
(1) & (2) & (3) & (4) & (1) & (2) & (3) & (4) & (1) & (2) & (3) & (4)\\
\hline
2212 & 2001-09-03 & 24.8 & 24.1 & 25934 & 2022-04-21 & 19.3 & 19.2 & 25969 & 2022-10-09 & 27.7 & 27.2\\
7712 & 2007-09-10 & 8.1 & 8.1 & 25935 & 2023-08-20 & 24.1 & 22.5 & 25970 & 2022-06-12 & 24.8 & 24.2\\
7915 & 2006-11-08 & 18.6 & 18.6 & 25936 & 2023-01-26 & 12.9 & 12.9 & 25971 & 2022-05-04 & 12.6 & 12.4\\
8477 & 2007-06-10 & 45.9 & 45.7 & 25937 & 2022-11-27 & 30.8 & 30.2 & 25972 & 2022-05-18 & 31.7 & 31.7\\
8557 & 2007-06-14 & 27.8 & 27.8 & 25938 & 2022-11-26 & 18.7 & 18.1 & 25973 & 2022-11-11 & 18.1 & 17.4\\
25277 & 2023-10-02 & 18.7 & 18.2 & 25939 & 2023-01-28 & 14.3 & 13.4 & 26280 & 2022-01-18 & 11.7 & 11.5\\
25278 & 2022-12-02 & 9.8 & 9.8 & 25940 & 2023-08-10 & 27.7 & 27.2 & 27347 & 2022-09-09 & 22.0 & 20.7\\
25279 & 2022-09-06 & 24.5 & 23.4 & 25941 & 2023-06-09 & 32.6 & 31.4 & 27449 & 2022-09-24 & 9.8 & 9.8\\
25907 & 2022-11-08 & 36.8 & 36.5 & 25942 & 2022-05-04 & 15.2 & 14.4 & 27450 & 2022-09-26 & 9.8 & 9.8\\
25908 & 2022-09-23 & 22.6 & 21.2 & 25943 & 2023-08-31 & 16.7 & 16.2 & 27556 & 2022-11-15 & 25.2 & 24.4\\
25909 & 2023-05-24 & 19.3 & 18.8 & 25944 & 2022-09-08 & 21.6 & 21.1 & 27563 & 2023-06-08 & 11.7 & 10.9\\
25910 & 2022-09-25 & 19.3 & 19.2 & 25945 & 2022-09-27 & 17.0 & 16.8 & 27575 & 2022-12-02 & 19.7 & 19.7\\
25911 & 2022-04-19 & 16.9 & 15.8 & 25946 & 2023-07-01 & 29.7 & 29.1 & 27678 & 2023-01-27 & 12.4 & 12.4\\
25912 & 2022-04-18 & 15.4 & 15.1 & 25947 & 2023-09-24 & 14.9 & 13.6 & 27679 & 2023-01-28 & 11.9 & 7.1\\
25913 & 2022-09-03 & 19.6 & 19.6 & 25948 & 2022-09-30 & 27.9 & 27.3 & 27680 & 2023-01-28 & 13.2 & 12.4\\
25914 & 2022-10-15 & 28.2 & 27.0 & 25949 & 2023-10-27 & 20.8 & 20.2 & 27681 & 2023-01-29 & 9.8 & 9.2\\
25915 & 2022-09-03 & 21.1 & 20.3 & 25950 & 2023-06-30 & 29.7 & 28.9 & 27739 & 2023-10-01 & 21.3 & 20.8\\
25916 & 2023-09-03 & 22.2 & 21.7 & 25951 & 2022-11-18 & 28.7 & 27.9 & 27780 & 2023-08-21 & 14.9 & 14.2\\
25917 & 2023-06-22 & 35.6 & 33.8 & 25952 & 2023-09-27 & 10.8 & 9.8 & 27856 & 2023-05-25 & 15.9 & 15.2\\
25918 & 2022-09-13 & 20.6 & 20.6 & 25953 & 2022-09-17 & 24.8 & 24.2 & 27857 & 2023-05-26 & 12.9 & 12.7\\
25919 & 2022-06-13 & 25.3 & 24.8 & 25954 & 2022-04-24 & 13.4 & 13.4 & 27896 & 2023-06-10 & 13.7 & 13.7\\
25920 & 2022-06-13 & 30.5 & 29.9 & 25955 & 2023-07-20 & 43.4 & 40.4 & 27974 & 2023-08-05 & 28.7 & 26.9\\
25921 & 2023-08-04 & 16.9 & 16.9 & 25956 & 2022-09-02 & 13.9 & 13.6 & 28370 & 2023-08-13 & 20.7 & 19.7\\
25922 & 2022-06-14 & 31.4 & 30.9 & 25957 & 2022-09-08 & 21.8 & 21.5 & 28483 & 2023-08-19 & 20.2 & 18.4\\
25923 & 2022-09-04 & 10.9 & 10.4 & 25958 & 2022-05-04 & 12.3 & 12.1 & 28872 & 2023-09-01 & 13.1 & 12.1\\
25924 & 2022-09-07 & 21.8 & 21.5 & 25959 & 2023-08-05 & 15.4 & 15.1 & 28886 & 2023-09-10 & 10.0 & 9.3\\
25925 & 2022-09-02 & 23.6 & 23.2 & 25960 & 2023-07-08 & 24.8 & 23.2 & 28887 & 2023-09-10 & 19.8 & 19.1\\
25926 & 2023-07-12 & 61.2 & 59.9 & 25961 & 2023-09-09 & 18.8 & 18.2 & 28910 & 2023-10-25 & 25.8 & 24.7\\
25927 & 2023-09-16 & 20.5 & 19.4 & 25962 & 2023-09-11 & 21.8 & 21.3 & 28920 & 2023-09-25 & 15.3 & 14.5\\
25928 & 2022-05-03 & 15.9 & 15.6 & 25963 & 2022-11-26 & 37.6 & 35.7 & 28934 & 2023-09-29 & 19.8 & 18.8\\
25929 & 2022-08-26 & 27.7 & 26.4 & 25964 & 2023-09-05 & 20.3 & 19.0 & 28951 & 2023-10-05 & 12.9 & 12.1\\
25930 & 2022-11-15 & 19.9 & 19.7 & 25965 & 2023-07-07 & 35.7 & 34.1 & 28952 & 2023-10-08 & 13.8 & 13.5\\
25931 & 2022-04-23 & 14.6 & 14.3 & 25966 & 2023-08-13 & 18.8 & 18.6 & 29207 & 2024-05-29 & 19.7 & 15.2\\
25932 & 2022-05-05 & 14.1 & 14.1 & 25967 & 2022-08-01 & 33.6 & 33.1 & 29427 & 2024-05-29 & 18.7 & 17.4\\
25933 & 2023-08-15 & 23.9 & 23.1 & 25968 & 2022-07-12 & 27.5 & 26.7 & \textbf{Total} & --- & \textbf{2200.6} & \textbf{2125.0}\\
\hline\hline
\end{tabular}
\begin{tablenotes}
\item
\textit{Notes.} Columns are: (1) Chandra observation ID, in ascending order; 
(2) observation start date; (3) total exposure time in ks; (4) cleaned exposure 
time in ks. The first observation (ObsID~=~2212) was carried out with ACIS-S; 
all remaining observations were conducted with ACIS-I.
\end{tablenotes}
\end{threeparttable}
\end{table*}
% ------------------------------------------------------------
\subsubsection{Observations and Data Reduction}
\label{sec: cxo_reduction}
This paper employs a list of Chandra datasets, obtained by the Chandra \mbox{X-ray} Observatory, contained in the Chandra Data Collection \dataset[DOI:10.25574/cdc.633]{https://doi.org/10.25574/cdc.633}. We consider two datasets: the same 1.25~Ms observations
used by \citet{Bogdan2024} (hereafter the ``B24 subset'') and an extended
2.2~Ms dataset that also includes 0.95~Ms of new observations obtained
between 2023 May and 2024 May (PI: A. Bogd\'an). Table~\ref{tbl: chandra_obs} reports the
utilized Chandra observations. UHZ1 was observed at a mildly off-axis angle, with an exposure-weighted mean off-axis angle of $1.9'$. The data were reduced with \texttt{CIAO}~4.18
and \texttt{CALDB}~4.12.3.\par

We first run the \texttt{chandra\_repro} script with the option \texttt{check\_vf\_pha = yes} because all the observations were taken in the Very Faint mode. We then conduct the first-pass merging of these observations with \texttt{merge\_obs} and use \texttt{wavdetect} to detect sources on the merged data between $0.5-7$~keV with a ``$\sqrt{2}$ sequence'' of wavelet scales (1, 1.414, 2, 2.828, 4, 5.656, and 8 pixels) and a significance threshold of $10^{-6}$. To assess any likely background flares, we then generate a source-free region by masking these detected source regions, where their sizes are enlarged by a factor of 2 and floored at $3''$, as well as the central $5'$ radius of Abell~2744. We then use \texttt{deflare} to clean background flares in each observation, leading to a $3.4\%$ reduction of the effective exposure time. This cleaning procedure generally has negligible effects. When only analyzing the B24 subset, we do not apply the deflare procedure to be consistent with \citet{Bogdan2024}.

A key systematic in this analysis is the relative
astrometric alignment of individual observations, which becomes important for
a marginal detection in a structured background. We therefore test three
astrometry correction methods before performing the final merge:
\begin{enumerate}
\item No astrometry corrections.
\item Same as \citet{Bogdan2024}, registration to reference X-ray sources detected in the single longest Chandra observation, i.e., ObsID = 8477 when analyzing the B24 subset or ObsID = 25926 when analyzing all the data.
\item Registration to reference optical sources in the Legacy Survey DR10 \citep{Dey2019}. To construct this reference list, we rely on the detected X-ray sources on the first-pass merged data. The \texttt{wavdetect} routine may return more spurious sources superimposed on the extended ICM emission (e.g., see Figure~10 in \citealt{Evans2024}). Therefore, to ensure reliable detections of compact sources, we also calculate their binomial no-source probability $P_B$ (\citealt{Weisskopf2007}; see also Equation~\ref{eq: PB} later) and only keep those above $5\sigma$. This cut primarily filters out sources located within the strong structured ICM. We then select bright sources with 100 net counts and 90\% PSF sizes $\leq2.5''$. These sources are then matched with the optical catalog with a matching radius of $1''$, and the matched optical positions are used as the reference list.
\end{enumerate}

When applicable, we use \texttt{fine\_astro} to apply astrometric corrections to the cleaned observations before performing a final merge. The impact of the above choices on the detection significance is quantified in Sections~\ref{sec: cxo_photometry} and~\ref{sec: cxo_astro}.\par
To assess the resulting astrometry of these astrometrically aligned and merged data, we detect \mbox{X-ray} sources with \texttt{wavdetect} again and further remove sources with $P_B$ below $5\sigma$. We select sources within $4'$ around UHZ1 and match them with the Legacy Survey DR10 and Gaia DR3 \citep{GaiaCollaboration2023} with a matching radius of $1.5''$ to obtain the corresponding optical counterparts. There are $\sim30$ sources matched to the Legacy Survey and 4 sources matched to Gaia. We report the key statistics of the offsets between the \mbox{X-ray} and optical position in Table~\ref{tbl: chandra_detection}, including the median, mean, the dispersion metric characterized by normalized median absolute deviation ($\sigma_\mathrm{NMAD}$), and standard deviation ($\sigma_\mathrm{std}$). Compared with the median and $\sigma_\mathrm{NMAD}$, the mean and $\sigma_\mathrm{std}$ are generally larger because they are more prone to outliers, and the offset distribution is asymmetric. Their differences thus reflect the typical range when summarizing the offsets with single-number statistics. The Gaia matching results are generally better because the corresponding \mbox{X-ray} sources are brighter and thus have better \mbox{X-ray} positions. As Table~\ref{tbl: chandra_detection} shows, compared with the first astrometry correction method (i.e., no correction), all the reported four statistics describing the astrometry generally improve after applying our third astrometry correction method (i.e., matching to optical positions). The latter results therefore show generally good consistency with the optical positions, indicating reliable final astrometry. The second astrometry correction (i.e., matching to \mbox{X-ray} positions) still improves the offset dispersion but behaves worse for the median or mean offsets. Especially, the ``Full (2)'' case shows a relatively large ($1''$) offset with the optical positions. This is because the specific chosen reference Chandra observations themselves have systematic offsets with optical positions. \citet{Bogdan2024} reported a mean offset of $0.33''$ and a dispersion of $0.14''$, which is similar to our ``B24 (2)'' result in Table~\ref{tbl: chandra_detection}, although the precise numbers are not directly comparable because they may have used different sources or methods when reporting the numbers.

% ------------------------------------------------------------
\subsubsection{Aperture Photometry and Detection Significance}
\label{sec: cxo_photometry}
% ------------------------------------------------------------

With the reduced and merged event files in hand, we perform
aperture photometry on the hard ($2$--$7$~keV), soft ($0.5$--$2$~keV), and
full ($0.5$--$7$~keV) bands, following the exact aperture configuration of
\citet{Bogdan2024} to enable a direct comparison. Specifically, we adopt a
circular source aperture of radius $r_{\rm src}=1\arcsec$ and a local
background annulus with radii $r_{\rm in}=3\arcsec$ and $r_{\rm out}=6\arcsec$
(area ratio $A_{\rm bkg}/A_{\rm src}=27$), both centered at R.A.\ =
00:14:16.096 and Dec.\ = $-$30:22:40.285. This source aperture corresponds to an encircled energy fraction of 0.8 at the nominal effective energy (3.8~keV) of the hard band.

Because UHZ1 lies in projection near the core of Abell~2744,
the local background is dominated by the hot ($kT\sim11$~keV) ICM and is substantially elevated relative to a blank field. Spatial
variations in the ICM surface brightness on arcsecond scales could in principle
bias the local background estimate, so we perform an independent check.
We model the $1'\times1'$ region around
UHZ1 (masking any detected point sources in addition to the UHZ1 aperture) with a 4th-order
polynomial in detector coordinates using Poisson statistics. The resulting
expected background in the $r_{\rm src}=1\arcsec$ aperture is consistent with the annular estimate for the same merged dataset at a $\lesssim 0.1$
count level, indicating that the mean
local background level is not strongly method-dependent for this test. Similar conclusions were also drawn by
\citet{Bogdan2024}. Nevertheless, sub-arcsecond variations in
the ICM surface brightness cannot be fully excluded, as they approach the
Chandra pixel scale.

We quantify the detection significance via the binomial
no-source probability $P_B$ \citep{Weisskopf2007}, which is better suited than a simple
signal-to-noise ratio for the low-count regime encountered here. We calculate:
\begin{align}
P_B=\sum_{X=S}^N\frac{N!}{X!(N-X)!}p^X(1-p)^{N-X}=I_\beta(p, S, N-S+1),\label{eq: PB}
\end{align}
where $S$ is the number of counts in the source region, $N$ is the number of
total counts in both the source and background regions,
$p=A_{\rm src}/(A_{\rm src}+A_{\rm bkg})$, and $I_\beta$ is the regularized incomplete
Beta function, which is useful for better numerical stability here. $P_B$ is
then converted to a nominal Gaussian significance level $Z_\mathrm{det}$ as follows:
\begin{align}
Z_\mathrm{det}=\sqrt{2}~\mathrm{erf^{-1}}(1-2P_B),\label{eq: Zdet}
\end{align}
where $\mathrm{erf^{-1}}$ is the inverse error function.\par

The results for all dataset and astrometry choices are
summarized in Table~\ref{tbl: chandra_detection}. Across all cases, we find
only a marginal signal in the hard band at the $2.0$--$2.9\sigma$ level,
substantially less significant than the $4.2$--$4.4\sigma$ reported by
\citet{Bogdan2024}. The soft band is fully undetected, and the full-band
significance reaches only $0.6-2.3\sigma$. Our lower detection significance has been cross-verified. Three authors of this work, all with expertise in \mbox{X-ray} astronomy, have independently analyzed the Chandra data and all obtained low detection significances. \citet{Li26} also independently reported their nondetection of UHZ1 (their Appendix~B). Finally, we note that UHZ1 was identified by \citet{Bogdan2024} among a sample of 11 $z>9$ candidates. If further considering the conservative Bonferroni correction and inflating $P_B$ by 11, the detection would be further undermined.

\begin{table*}
\begin{threeparttable}
\caption{Detection Significance Summary}
\label{tbl: chandra_detection}
\begin{tabular*}{\textwidth}{@{\extracolsep{\fill}}c|cc|ccc|ccc|ccc}
\hline\hline
Data & \multicolumn{2}{c|}{Offset} & \multicolumn{3}{c|}{Soft band} & \multicolumn{3}{c|}{Hard band} & \multicolumn{3}{c}{Full band}\\
\hline
& Median/Mean & $\sigma_\mathrm{NMAD}/\sigma_\mathrm{std}$ & $C_\mathrm{tot}$ & $C_\mathrm{net}$ & $Z_\mathrm{det}$ & $C_\mathrm{tot}$ & $C_\mathrm{net}$ & $Z_\mathrm{det}$ & $C_\mathrm{tot}$ & $C_\mathrm{net}$ & $Z_\mathrm{det}$\\
(1) & (2) & (3) & (4) & (5) & (6) & (4) & (5) & (6) & (4) & (5) & (6)\\
\hline
\citeauthor{Bogdan2024} & $0.33''$ & $0.14''$ & & & & 42 & 20.6 & $4.2-4.4\sigma$\\
(\citeyear{Bogdan2024}) & & & & & & & & $(4.0-4.3\sigma)$\\
\hline
B24 (1) & $0.24''/0.36''$ & $0.22''/0.29''$ & 19 & $-1.4$ & $-0.4\sigma$ & 34 & 14.5 & $2.8\sigma$ & 53 & 13.1 & $1.9\sigma$\\
& $0.16''/0.22''$ & $0.09''/0.19''$ & & & & & &\\
B24 (2) & $0.30''/0.36''$ & $0.11''/0.22''$ & 20 & $-0.6$ & $-0.2\sigma$ & 35 & 15.1 & $2.9\sigma$ & 55 & 14.6 & $2.1\sigma$\\
& $0.31''/0.30''$ & $0.07''/0.07''$ & & & & & &\\
B24 (3) & $0.20''/0.30''$ & $0.17''/0.28''$ & 14 & $-6.5$ & $-1.6\sigma$ & 33 & 13.3 & $2.6\sigma$ & 47 & 6.8 & $1.0\sigma$\\
& $0.20''/0.19''$ & $0.15''/0.12''$ & & & & & &\\
\hline
Full (1) & $0.38''/0.48''$ & $0.21''/0.31''$ & 34 & 1.8 & $0.2\sigma$ & 54 & 19.1 & $2.9\sigma$ & 88 & 20.9 & $2.3\sigma$\\
& $0.25''/0.24''$ & $0.05''/0.05''$ & & & & & &\\
Full (2) & $1.01''/1.01''$ & $0.18''/0.22''$ & 29 & $-1.5$ & $-0.3\sigma$ & 49 & 15.1 & $2.3\sigma$ & 78 & 13.6 & $1.6\sigma$\\
& $0.99''/1.00''$ & $0.03''/0.03''$ & & & & & &\\
Full (3) & $0.21''/0.33''$ & $0.18''/0.30''$ & 24 & $-7.4$ & $-1.4\sigma$ & 47 & 12.7 & $2.0\sigma$ & 71 & 5.3 & $0.6\sigma$\\
& $0.11''/0.10''$ & $0.03''/0.04''$ & & & & & &\\
\hline\hline
\end{tabular*}
\begin{tablenotes}
\item
\textit{Notes.} (1) Merged dataset used. The first two rows reproduce the values reported in \citet{Bogdan2024}. Its first row is their originally reported values, and the second row shows the corrected $Z_\mathrm{det}$ within parentheses based on Equation~\ref{eq: Zdet}. ``B24'' and ``Full'' denote the 1.25~Ms B24 subset and the 2.2~Ms full dataset, respectively; the number in parentheses indicates the astrometry correction method applied (see Section~\ref{sec: cxo_reduction}). (2) and (3) The median, mean, $\sigma_\mathrm{NMAD}$, and $\sigma_\mathrm{std}$ of the offsets between the \mbox{X-ray} sources and their optical counterparts. Each combination of the dataset and the adopted astrometry correction method has two rows of them, with the first representing the matching results with the Legacy Survey and the second for Gaia. (4) Total counts within the $1''$ source aperture. (5) Net counts within the $1''$ source aperture. (6) Detection significance.
\end{tablenotes}
\end{threeparttable}
\end{table*}

% ------------------------------------------------------------
\subsubsection{Sensitivity to Astrometric Registration}
\label{sec: cxo_astro}
% ------------------------------------------------------------

The discrepancy between the hard-band significance reported above
and the value reported by \citet{Bogdan2024}
motivates a quantitative assessment of how sensitive a
marginal detection in this field is to sub-arcsecond astrometric registration
choices. There is one subtle difference in the $Z_\mathrm{det}$ calculation (Equation~\ref{eq: Zdet}) between this work and \citet{Bogdan2024}, who adopted $Z_\mathrm{det}=\sqrt{2}~\mathrm{erf^{-1}}(1-P_B)$. This difference reflects that we regarded $P_B$ as a one-tail probability, while \citet{Bogdan2024} regarded it as two-tail. However, $P_B$, by definition, accounts for only the tail with source-region counts equal or larger than the observed ones, and thus our Equation~\ref{eq: Zdet} is more appropriate. Therefore, the significance in \citet{Bogdan2024} would slightly drop from $4.2-4.4\sigma$ to $4.0-4.3\sigma$ if adopting Equation~\ref{eq: Zdet}. Even after this modification, the discrepancy in the hard-band significance still exists.\par
We only had three main procedures from the raw Chandra data to counts measurements: data reprocessing (\texttt{chandra\_repro}), background flare cleaning (\texttt{deflare}), and astrometric correction (\texttt{fine\_astro}). The first two steps are rather conceptually straightforward and are unlikely to have material impacts. We tried skipping these two steps and directly counting photons from the archived Chandra level 1 and level 2 event files. Here ``level 1'' files contain all the uncleaned events, while ``level 2'' files have undergone standard pipeline cleaning (especially only keeping the standard ASCA grade set 02346) and hence are more science-ready. The level 1 (2) files return $C_\mathrm{tot}=57~(55)$, $C_\mathrm{net}=18.1~(18.9)$, and a significance of $2.6\sigma~(2.8\sigma)$ in the hard band. These are close to our reported ``Full (1)'' results in Table~\ref{tbl: chandra_detection}: $C_\mathrm{tot}=54$, $C_\mathrm{net}=19.1$, and $2.9\sigma$. Therefore, the remaining most likely reason for the discrepancy is the astrometric correction.\par
\begin{figure}
\includegraphics[width=\linewidth]{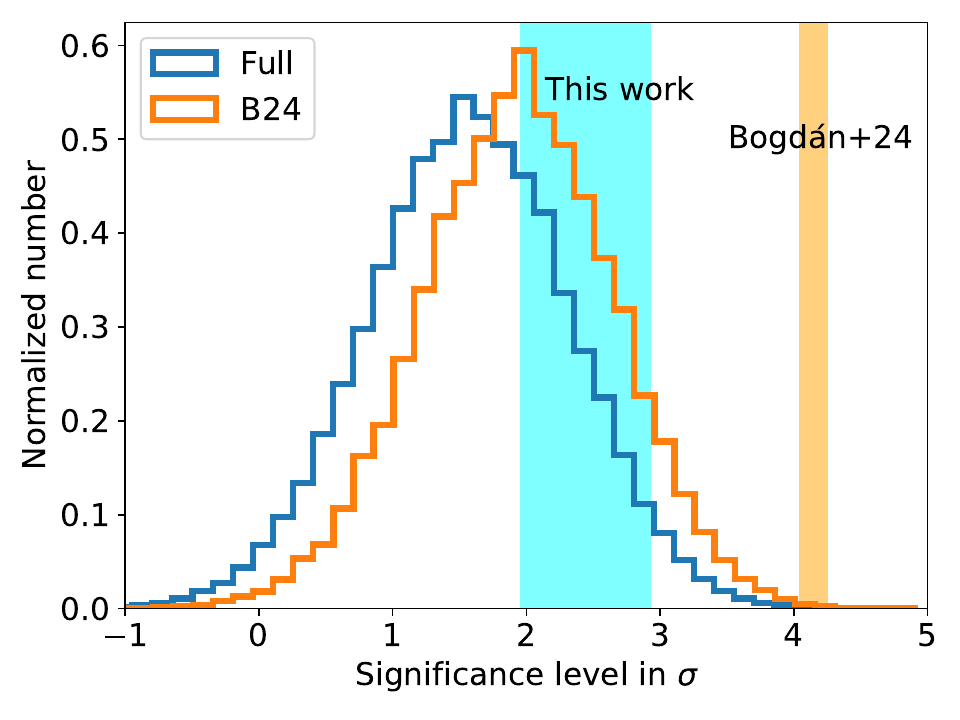}
\caption{Distribution of hard-band detection significance obtained by randomly 
shifting the astrometric registration of each observation. Both histograms are 
normalized to unit sum. Blue and orange histograms correspond to the full 
2.2~Ms dataset and the B24 subset \citep{Bogdan2024}, respectively. The cyan 
shaded band marks the range of detection significances measured in this work 
(Table~\ref{tbl: chandra_detection}); the orange shaded band marks the 
significance from \citet{Bogdan2024}. The $4.0-4.3\sigma$ value of 
\citet{Bogdan2024} is recovered in fewer than $0.1\%$ of Monte Carlo 
realizations, demonstrating its sensitivity to the assumed astrometric 
alignment.}
\label{fig: randomshift}
\end{figure}

We experiment by randomly shifting each observation. The shifting offsets are generated
independently from a Gaussian distribution with a $0.5''$ dispersion along
both the R.A.\ and Dec.\ directions, and thus the absolute offset would
follow a Rayleigh distribution with a 68\% quantile of $0.8''$ and a 90\%
quantile of $1.1''$, consistent with the expected Chandra absolute astrometric
accuracy.\footnote{\url{https://cxc.cfa.harvard.edu/cal/ASPECT/celmon/}} We
conduct this random shifting $10^5$ times and plot the distribution of the
resulting detection significance in Figure~\ref{fig: randomshift}.

The broad distribution in Figure~\ref{fig: randomshift}
demonstrates that astrometric uncertainties alone can introduce or suppress
apparent excesses of up to $3$--$4\sigma$ at this source position.
The $4.0-4.3\sigma$ of \citet{Bogdan2024}, after the modification in Equation~\ref{eq: Zdet}, resides in the
extreme tail of the distribution, with only $0.03\%$ ($0.1\%$) of the B24
(full) random-shifting results reaching $4.0\sigma$.
This suggests that the original detection is sensitive to a particularly favorable astrometric alignment, and that the significance is not robustly recovered across the full range of plausible 
reductions of the same data.

% ------------------------------------------------------------
\subsubsection{Temporal Behavior of the Hard-Band Signal}
\label{sec: cxo_lc}
% ------------------------------------------------------------
A genuine steady X-ray source should produce net counts that grow
proportionally with exposure time. We use the extended 2.2~Ms dataset to test
whether the hard-band signal at the UHZ1 position is consistent with a steady
source. When the dataset is extended from the 1.25~Ms B24 subset to a total
exposure of 2.2~Ms, the inferred net counts do not increase in
proportion to exposure time, and their source signal significance remains
similarly marginal in Table~\ref{tbl: chandra_detection}.

\begin{figure*}
\centering
\includegraphics[width=0.8\linewidth]{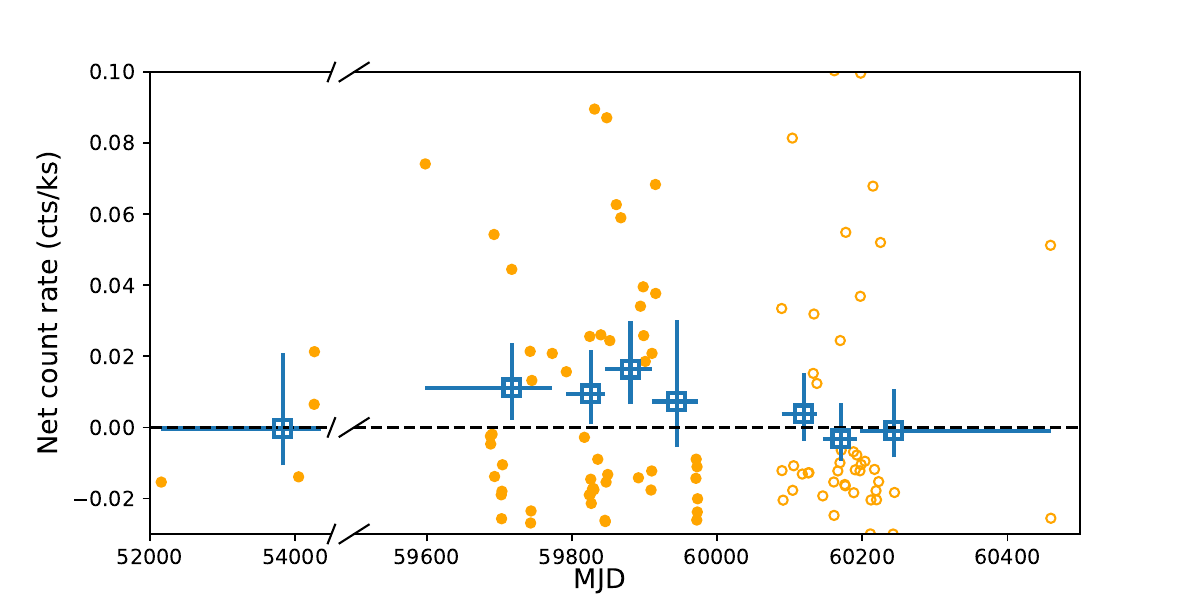}
\caption{Hard-band net count rates in the $1''$ source aperture. Each orange point stands for one observation, where the filled (open) ones represent those in (not in) the B24 subset. The open squares are binned to reach $\approx100-300$~ks. Their abscissa error bars represent the timespan of the corresponding bin, and their ordinate error bars reflect $1\sigma$ source-region count uncertainties calculated based on \citet{Gehrels1986}. The net count rate is consistent with zero in the more recent 0.95~Ms observations.}
\label{fig: lc}
\end{figure*}

To examine this behavior across individual epochs,
Figure~\ref{fig: lc} shows the hard-band net count rate in the $1''$ source
aperture for each individual observation based on the ``Full (3)'' dataset in Table~\ref{tbl: chandra_detection}. The binned light curve constructed
from individual observations shows a positive net count rate in earlier epochs,
but the net rate is consistent with zero in the more recent 0.95~Ms
observations. Considering the more recent 0.95~Ms data alone, their 90\% upper limit for the hard-band net count rate is $8.3~\mathrm{Ms^{-1}}$, calculated following \citet{Weisskopf2007}. If the net count rate remains constant, the expected net counts in the 1.25~Ms B24 subset should be $<10$, lower than the observed net counts ($\sim14$) in Table~\ref{tbl: chandra_detection}. However, this difference is not statistically sufficient to claim variability given the additional statistical uncertainty ($\sim6$ counts) of the net-count estimation in the 1.25~Ms B24 subset and the overall marginal detection significance.
Nevertheless, the failure of the signal to
accumulate as expected for a steady source further undermines the case for a
robust detection.

%----------------------------------------------------------------------
\subsection{JWST/MIRI Imaging}
\label{sec: miri}
%----------------------------------------------------------------------

We analyze JWST/MIRI imaging observations of UHZ1 obtained as part 
of PID~6123 (PIs: E. Gallo \& G. Roberts-Borsani). All the JWST data used in this paper can be found in MAST: \dataset[https://doi.org/10.17909/namr-v852]{https://doi.org/10.17909/namr-v852}. The data span the full suite of nine MIRI 
broadband filters, from F560W to F2550W, providing continuous coverage of the 
observed-frame mid-infrared between $5.6$ and $25.5\,\mu$m, corresponding to 
rest-frame wavelengths of approximately $0.5$--$2.3\,\mu$m at the redshift of 
UHZ1.

We retrieve the stage~3 mosaics (\texttt{\_i2d.fits}) from the Mikulski 
Archive for Space Telescopes (MAST). We mask detected sources and then estimate the background map using \texttt{photutils} \citep{Bradley2025}, with 
a $3\sigma$ clipping threshold to exclude potential sources. Visual inspection 
of each background-subtracted mosaic at the known position of UHZ1 reveals no 
source above the local noise level in any of the nine filters.

To quantify the significance of the non-detections, we extract the flux 
within a circular aperture enclosing $50\%$ of the PSF energy in each band, centered on the 
UHZ1 position. The local background RMS noise is estimated as the standard 
deviation of fluxes measured in apertures of identical radius, placed randomly 
in nearby source-free regions. The signal-to-noise ratio is consistent with 
zero in all nine bands, confirming that UHZ1 is undetected across all MIRI 
imaging filters. A variance-weighted stack of all nine background-subtracted 
mosaics likewise yields a non-detection.

\begin{table}
\caption{JWST/MIRI photometric results}
\label{tbl: miri}
\centering
\begin{threeparttable}
\begin{tabular*}{\hsize}{@{\extracolsep{\fill}}ccccc}
\hline\hline
Band & Wavelength & Exposure time & Radius & Flux\\
& ($\mu\mathrm{m}$) & (seconds) & & ($\mu\mathrm{Jy}$)\\
(1) & (2) & (3) & (4) & (5)\\
\hline
F560W & 5.64 & 6649 & $0.21''$ & $<0.13$\\
F770W & 7.64 & 3519 & $0.23''$ &$<0.23$\\
F1000W & 9.95 & 1987 & $0.25''$ &$<0.36$\\
F1130W & 11.31 & 4163 & $0.27''$ &$<0.56$\\
F1280W & 12.81 & 2231 & $0.29''$ &$<0.66$\\
F1500W & 15.06 & 1010 & $0.33''$ &$<1.14$\\
F1800W & 17.98 & 1232 & $0.39''$ &$<2.17$\\
F2100W & 20.80 & 2398 & $0.45''$ &$<6.07$\\
F2550W & 25.36 & 5483 & $0.51''$ &$<26.23$\\
\hline\hline
\end{tabular*}
\end{threeparttable}
\begin{tablenotes}
\item
\textit{Notes.} (1) JWST/MIRI imaging filter used for the observations. (2) Pivot wavelength of the broadband filter. (3) Total on-source exposure time for each filter. The exposure times vary among filters because the observations were optimized to achieve comparable sensitivity limits given the wavelength-dependent background and detector performance of MIRI. (4) The circular aperture radii enclosing 50\% the PSF energy, retrieved from the JWST Calibration References Data System. (5) Measured $3\sigma$ flux density upper limits derived from aperture photometry at the galaxy center after background subtraction and aperture correction.
\end{tablenotes}
\end{table}

The resulting $3\sigma$ upper limits on the mid-infrared flux density of UHZ1  are listed in Table~\ref{tbl: miri} and are used in Section~\ref{sec: discussion} to place upper limits on the luminosity of any buried AGN.

%--------------------------------
\section{Discussion}
\label{sec: discussion}
%--------------------------------
We present a reassessment of the evidence for a luminous, heavily obscured AGN in UHZ1 at z$\simeq$10.05, combining a reanalysis of the full 2.2~Ms Chandra dataset with new JWST/MIRI photometric limits on the mid-infrared continuum. Our principal conclusion is that no compelling evidence for such an AGN exists. The case for one rests on a hard X-ray excess that, under closer scrutiny, proves to be of marginal statistical significance. Our reanalysis of the same 1.25~Ms Chandra dataset used by \citet{Bogdan2024}, and of the extended 2.2~Ms
dataset incorporating all available observations through 2024 May, fails to
reproduce a robust detection. Across reasonable, reproducible reductions of
the $2$--$7$~keV data, the inferred detection significance at the UHZ1
position spans $\sim2.0$--$2.9\sigma$ and is sensitive to sub-arcsecond
astrometric registration (Table~\ref{tbl: chandra_detection};
Section~\ref{sec: cxo_photometry}). The Monte Carlo
astrometric analysis (Section~\ref{sec: cxo_astro}) shows that the
$4.2$--$4.4\sigma$ ($4.0-4.3\sigma$ after modification based on Equation~\ref{eq: Zdet}) reported by \citet{Bogdan2024} lies in the extreme tail of
the distribution of outcomes achievable under plausible registration choices,
with only $0.1\%$ of random realizations reaching that significance.
Moreover, when the dataset is extended from 1.25~Ms to 2.2~Ms, the hard-band
net counts do not grow in proportion to the added exposure time: the 0.95~Ms
of new observations contribute negligible additional signals
(Section~\ref{sec: cxo_lc}), behavior that is inconsistent with a steady
point source. Taken together, the evidence for a point-like X-ray counterpart to UHZ1 is not statistically robust.\\

\begin{figure*}
\centering
\includegraphics[width=0.75\linewidth]{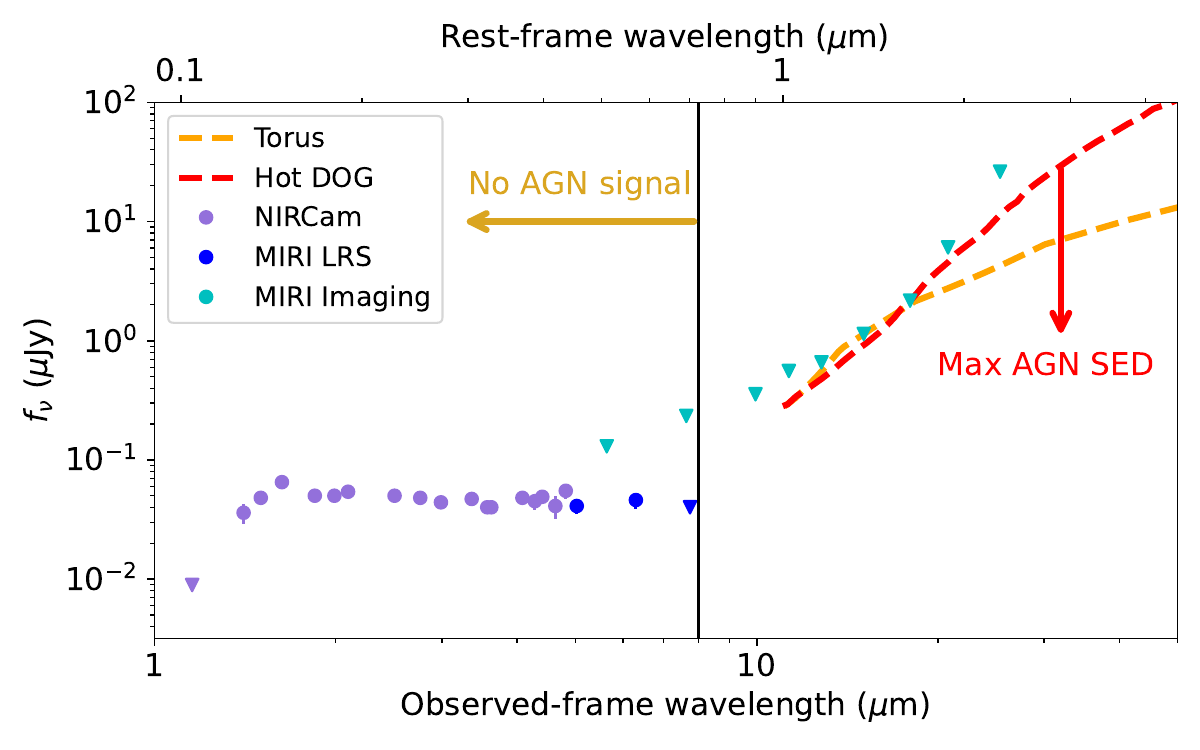}
\caption{Observed SED of UHZ1. Filled circles are photometric detections 
and downward triangles are $3\sigma$ upper limits. The vertical black line 
marks $\lambda_{\rm obs}=8~\mu\mathrm{m}$ (rest-frame $\sim\!0.7~\mu\mathrm{m}$), 
separating the NIRCam and MIRI LRS from the redder MIRI imaging bands. At 
$\lambda_{\rm obs}<8~\mu\mathrm{m}$, the SED is fully consistent with an 
unobscured, low-mass, metal-poor, star-forming galaxy with no AGN contribution. At $\lambda_{\rm obs}>8~\mu\mathrm{m}$, the MIRI non-detections place a direct upper bound on the bolometric luminosity of any 
buried AGN. The dashed orange and red curves show the maximally allowed 
normalization of the Torus \citep{Polletta2006} and Hot DOG \citep{Fan2016} AGN 
SED templates, respectively, each scaled to the F1500W and F1800W upper limits in Table~\ref{tbl: miri}. The
bolometric luminosity implied by these upper envelopes falls one to three
orders of magnitude below the intrinsic X-ray luminosity inferred by
\citet{Bogdan2024} under a Compton-thick model, ruling out a luminous
buried AGN of that magnitude in UHZ1.}
\label{fig: sed}
\end{figure*}

The SED of UHZ1 is shown in Figure~\ref{fig: sed}, with data from JWST/NIRCam \citep{Suess2024}, JWST/MIRI LRS continuum 
\citep{Alvarez-Marquez2026}, and the MIRI photometric limits derived in this work (Section~\ref{sec: miri}). 
Independent of the X-ray analysis, the JWST/MIRI photometry at $\lambda_{\rm obs}>8~\mu\mathrm{m}$ (Section~\ref{sec: miri}) yields an upper limit to the bolometric luminosity of any AGN that may be buried within  
UHZ1. 
To quantify this 
limit, we consider two AGN SED templates: 
(\textit{i}) the ``Torus'' template of \citet{Polletta2006}, and (\textit{ii}) 
the extremely red SED of hot dust-obscured galaxies \citep[Hot DOGs;][]{Fan2016}.
The former is for typical heavily obscured quasars, while the latter represents the most conservative benchmark case because Hot DOGs are among the reddest, most 
infrared-luminous galaxies known at $z\lesssim4$ and are powered by deeply 
buried, rapidly accreting AGN. We restrict the template fits to rest-frame 
wavelengths $>1~\mu\mathrm{m}$ (i.e., observed-frame $>11~\mu\mathrm{m}$), 
since shorter-wavelength emission is (\textit{i}) more susceptible to 
host-galaxy obscuration,\footnote{Nevertheless, the host galaxy of UHZ1 has 
only little obscuration, with an SED-based best-fit $A_V=0.2$~mag \citep{Alvarez-Marquez2026}.} and (\textit{ii}) subject to either a highly 
uncertain leaking fraction of intrinsic optical light or non-negligible 
host-galaxy contamination. Each template is scaled upward until it reaches the tightest MIRI
constraint: F1500W for the Torus template and F1800W for the Hot DOG
template, as shown in Figure~\ref{fig: sed}.\par

After correcting for a gravitational magnification factor of
$3.71_{-0.23}^{+0.23}$ \citep{Bergamini2023}, the MIRI non-detections constrain the
bolometric luminosity of any buried AGN. For the Torus template we
obtain $L_{\rm bol} < 1.3 \times 10^{45}$~erg~s$^{-1}$. Applying the
same $2$--$10$~keV bolometric correction as \citet{Bogdan2024}
($L_{\rm bol}/L_{\rm X,int} = 21$--$73$), this limit translates into
an upper bound to the intrinsic X-ray luminosity of $L_{\rm X,int} <
(1.8$--$6.2) \times 10^{43}$~erg~s$^{-1}$. For the Hot DOG template, the far-infrared emission may carry a
non-negligible contribution from host starburst activity \citep{Tsai2015},
making a clean bolometric inference unreliable. 
Separately, we use both templates to constrain the rest-frame $6~\mu$m AGN luminosity,
$L_{6~\mu\mathrm{m}}$, which is tightly correlated with $L_{\rm X}$ and
largely free from host contamination.
We obtain $L_{6~\mu\mathrm{m}} < 
3.1\times10^{44}~\mathrm{erg~s^{-1}}$ and $L_{6~\mu\mathrm{m}} < 3.0\times10^{45}~\mathrm{erg~s^{-1}}$ for the Torus and Hot DOG templates, 
respectively. 
Using the $L_{6\,\mu\mathrm{m}}$--$L_{\rm X,int}$ relation of
\citet{Stern2015}, these yield $L_{\rm X,int} < 9.6 \times 10^{43}$
and $< 4.2 \times 10^{44}$~erg~s$^{-1}$, respectively. As an additional check, we also use the rest-frame $K$-band ($2.1~\mu\mathrm{m}$) upper limit, interpolated between the F2100W and F2550W upper limits, to infer the $L_{\rm X,int}$ limit based on the observed local Seyfert-galaxy correlations (extrapolated toward high luminosities) in \citet{Burtscher2015} and \citet{Muller-Sanchez2018}. We obtain $L_{\rm X,int}<2\times10^{45}-8\times10^{46}~\mathrm{erg~s^{-1}}$, with exact values depending upon whether the Seyfert~1 or 2 relation is used since Seyfert~1 AGNs are much brighter in the near-IR at a fixed $L_{\rm X,int}$ \citep{Burtscher2015}. This $K$-band-based upper limit is not in tension with our SED-based constraints but is much looser because the F2100W and F2550W upper limits are higher than for F1500W and F1800W and because the local relations must be extrapolated well beyond their calibrated luminosity range. We thus still rely on the SED-based $L_{\rm X,int}$ limits. Combining all
$L_{\rm bol}$- and $L_{6\,\mu\mathrm{m}}$-based constraints, the MIRI
data limit the intrinsic X-ray luminosity of any buried AGN in UHZ1 to
$L_{\rm X,int} < (1.8$--$42) \times 10^{43}$~erg~s$^{-1}$---one to
three orders of magnitude below the value inferred by \citet{Bogdan2024}
under a Compton-thick model ($L_{\rm X,int} = 9 \times 10^{45}$~erg~s$^{-1}$,
$N_{\rm H} = 8\times 10^{24}$~cm$^{-2}$), itself
uncertain given the low significance of the X-ray detection reported
in our reanalysis.\par
The analyses above adopt SED templates dominated by hot dust emission. There are two possibilities that can suppress the dust emission in the MIRI bands: (\textit{i}) the obscuring gas in UHZ1 is dust-free, which is plausible in the early Universe, and no IR dust emission would then be expected; and (\textit{ii}) the obscuring medium is far from the central AGN and thus cooler, and the dust would emit mainly in the far-IR. However, these two possibilities are inconsistent with other multiwavelength data in the literature. In the first possibility, the rest-frame UV-to-optical AGN SED would remain visible (e.g., \citealt{Natarajan2024,Pacucci2026}), and the NIRCam data in Figure~\ref{fig: sed} would impose considerably stronger constraints. The upper limits derived from the MIRI non-detections under the hot-dust assumption are therefore more conservative. Regarding the second possibility, Section~4.1 in \citet{Algera2025} has explicitly discussed and ruled out this scenario. Briefly, their inferred dust mass is $\lesssim5.5\times10^5~M_\odot$ based on the nondetection of the far-IR dust continuum by the Atacama Large Millimeter/submillimeter Array (ALMA), and the low optical attenuation $A_V$ also suggests that the dust reservoir in the larger-scale interstellar medium is limited. Therefore, the host galaxy itself is unable to provide an obscuration at a Compton-thick level, and any such obscuration must be close to the nucleus.\par
The X-ray and mid-IR results presented here are corroborated
by a consistent body of evidence from independent JWST observations, which we
summarize for completeness. In each case, the data disfavor the presence of
a luminous obscured AGN:
\begin{enumerate}
    \item Absence of AGN-like spectral features in the rest-UV
    \citep{Goulding2023}: the NIRSpec spectrum lacks the prominent
    high-ionization UV lines typically associated with accreting sources at
    high redshift (e.g., strong \iona{C}{iv}, \iona{N}{v}, \iona{He}{ii}), in
    contrast to confirmed $z\gtrsim9$ systems with comparable claimed X-ray
    luminosities \citep{Kovacs2024,Napolitano2025}.

    \item Low-excitation rest-optical nebular spectrum
    \citep{Alvarez-Marquez2026}: MIRI/LRS detects
    H$\beta$+[\iona{O}{iii}] and H$\alpha$ with unusually low excitation
    compared to both star-forming and AGN samples at $z\gtrsim3$. The
    inferred ionization parameter ($\log U \simeq -2.5$) and optical
    diagnostic ratios occupy the locus of low-metallicity stellar
    photoionization models rather than AGN grids.

    \item No evidence for the large dust obscuration required by a
    Compton-thick interpretation \citep{Algera2025, Alvarez-Marquez2026}:
    Balmer decrements are consistent with Case~B recombination
    ($A_V \lesssim 0.4$\,mag). Deep ALMA non-detections independently
    constrain the cold dust mass to $M_{\rm dust}\lesssim
    5.5\times10^5\,M_\odot$, providing little support for a dust-rich host
    capable of hiding a luminous nucleus. Their ALMA and our MIRI non-detections suggest that there is no significant (either hot or cold) dust re-emission that would be expected from absorbing powerful AGN lights.

    \item Tension between recombination-line luminosity and claimed
    X-ray power \citep{Alvarez-Marquez2026}: standard
    $L_{\mathrm{H}\alpha}$--$L_\mathrm{X}$ scalings predict
    $L_{\rm 2-10\,keV}\sim\mathrm{few}\times10^{43}$\,erg\,s$^{-1}$ from
    the observed H$\alpha$ luminosity, at least an order of magnitude below
    the Compton-thick X-ray inference, with the absence of high-excitation
    narrow lines further weakening the type~2 AGN hypothesis.
\end{enumerate}

These results have direct implications for the interpretation of UHZ1
as a direct collapse black hole candidate. The high black hole mass
($\sim 10^7$--$10^8\,M_\odot$) inferred from the X-ray signal by
\citet{Bogdan2024} was taken to imply a DCBH seed, given the limited
time available for black hole growth at $z \simeq 10$. The DCBH
scenario further requires an outsize black hole---with $M_{\rm BH}$
comparable to or exceeding $M_\star$---that dominates the bolometric
output, producing a source luminous in hard X-rays yet faint in the
rest-frame UV--optical \citep{Natarajan2017,Pacucci2023}.
The marginal significance of the X-ray detection undermines the claim
of a luminous buried AGN; the mid-IR SED places a stringent upper
bound on any such luminosity; the rest-frame UV--optical is consistent
with stellar photoionization at low metallicity; and the dust content
is insufficient to support the Compton-thick column density required by
the \citet{Bogdan2024} spectral model. None of the multiwavelength
criteria expected of a DCBH candidate are satisfied beyond
reasonable doubt.
More broadly, this case illustrates the risks of inferring
BH masses from marginal X-ray detections in cluster fields where the ICM provides a bright, spatially structured background.
The long inference chain from raw counts to $M_{\rm BH}\sim10^7$--$10^8\,M_\odot$ can propagate and amplify systematic errors to produce apparently compelling results that do not survive independent scrutiny. 

Given that the claim of a DCBH seed at $z\simeq10$ is extraordinary, the  burden of proof is correspondingly high. A comprehensive reanalysis of all available
2.2~Ms Chandra data, combined with new JWST/MIRI photometric limits
and existing JWST spectroscopy, finds no compelling evidence for a luminous, 
obscured AGN in UHZ1 at $z\simeq10.05$. The multiwavelength data are 
consistent with UHZ1 being a low-mass ($M_\star\sim10^8\,M_\odot$),
metal-poor ($Z\sim0.04\,Z_\odot$), star-forming
($\mathrm{SFR}\sim1\,M_\odot\,\mathrm{yr}^{-1}$) galaxy.

% ------------------------------------------------
\begin{acknowledgements}
This work is based in part on observations made with the NASA/ESA/CSA James
Webb Space Telescope. The data were obtained from the Mikulski Archive for
Space Telescopes at the Space Telescope Science Institute, which is operated
by the Association of Universities for Research in Astronomy, Inc., under
NASA contract NAS 5-03127 for JWST; these observations are associated with
program \#6123. Support for program \#6123 was provided by NASA through a
grant from the Space Telescope Science Institute under the same contract.
The scientific results reported in this article are also based in part on
data obtained from the Chandra Data Archive. The authors thank the anonymous referee for constructive comments and suggestions.
\end{acknowledgements}

\bibliographystyle{aasjournalv7.1}
\bibliography{bibliography.bib}

\end{document}